\documentclass[9.5pt, conference]{IEEEtran}
\pdfoutput=1
\IEEEoverridecommandlockouts
\usepackage{pdfpages}
\usepackage{cite}
\usepackage{amsmath,amssymb,amsfonts}
\usepackage{algorithmic}
\usepackage{graphicx}
\usepackage{textcomp}
\usepackage{xcolor}

\usepackage{amssymb}
\DeclareMathOperator{\EX}{\mathbb{E}}

\usepackage{todonotes}

\usepackage[utf8]{inputenc}
\usepackage[T1]{fontenc}
\usepackage{enumitem}

\def\BibTeX{{\rm B\kern-.05em{\sc i\kern-.025em b}\kern-.08em
    T\kern-.1667em\lower.7ex\hbox{E}\kern-.125emX}}
   
\begin{document}
\bibliographystyle{IEEEtran}

\title{End-to-End Conditional GAN-based Architectures for Image Colourisation \\
\thanks{The work described in this paper has been conducted within the project JOLT. This project is funded by the European Union’s Horizon 2020 research and innovation programme under the Marie Skłodowska Curie grant agreement No 765140.}
}

\author{\IEEEauthorblockN{Marc Górriz Blanch, Marta Mrak}
\IEEEauthorblockA{\textit{BBC Research \& Development, London, UK} \\
marc.gorrizblanch@bbc.co.uk \\
}

\and
\IEEEauthorblockN{Alan F. Smeaton, Noel E. O'Connor}
\IEEEauthorblockA{\textit{Dublin City University} \\
Glasnevin, Dublin, Ireland\\ 
}
}


\IEEEaftertitletext{\vspace{-1.8\baselineskip}}
\maketitle

\begin{figure*}[htbp]
\vspace{-2\baselineskip}
\centerline{\includegraphics[width=1\textwidth]{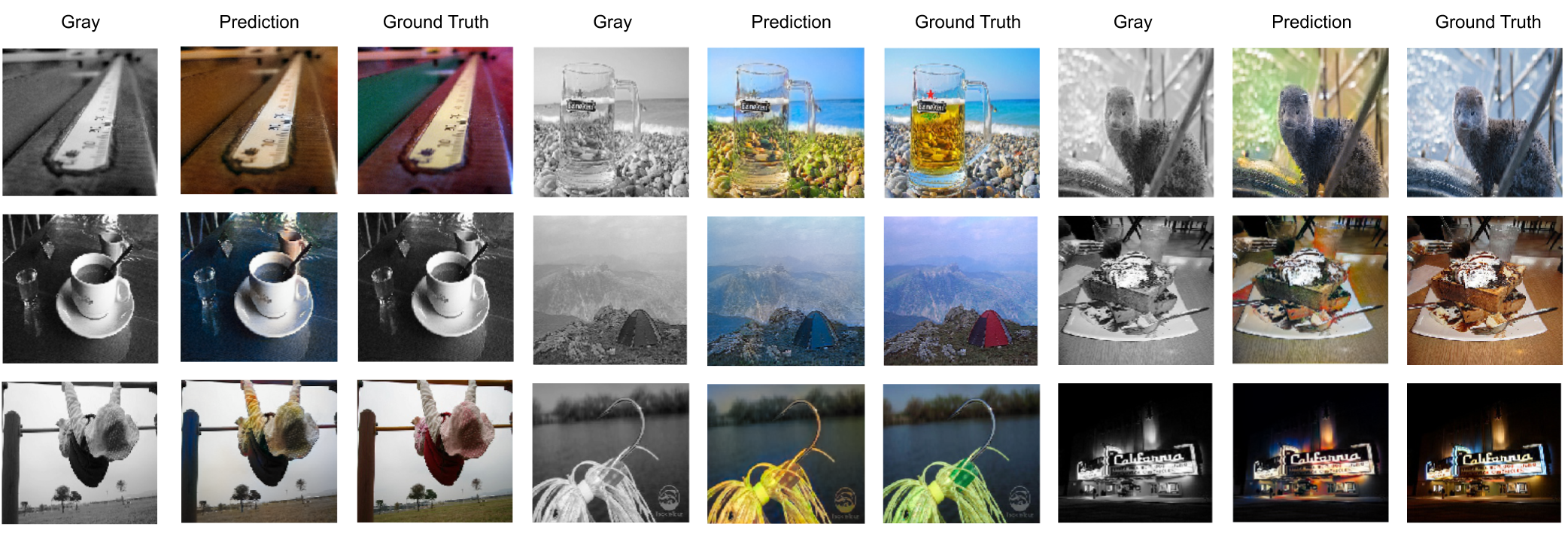}}
\vspace{-0.2\baselineskip}
\caption{Visualisation of  colourisation performance with the presented  architecture mixing batch and instance normalisation.}
\label{fig1}
\vspace{-0.5\baselineskip}
\end{figure*}

\begin{abstract}
In this work  recent advances in conditional adversarial networks are investigated to develop an end-to-end architecture based on Convolutional Neural Networks (CNNs) to directly map realistic colours to an input greyscale image. Observing that  existing colourisation methods sometimes exhibit a lack of colourfulness, this paper proposes a method to improve colourisation results. In particular, the  method uses Generative Adversarial Neural Networks (GANs) and focuses on improvement of training stability to enable better generalisation in large multi-class image datasets. Additionally, the integration of instance and batch normalisation layers in both generator and discriminator is introduced to the popular U-Net architecture, boosting the network capabilities to generalise the style changes of the content. The  method has been tested using the ILSVRC 2012 dataset, achieving improved automatic colourisation results compared to other methods based on GANs.

\end{abstract}

\begin{IEEEkeywords}
Colourisation, Conditional GANs, CNNs.
\end{IEEEkeywords}

\section{Introduction}
Colourisation refers to the process of adding colours to greyscale or other monochrome images such that the colourised results are perceptually meaningful and visually appealing. In general, greyscale content is present in many multimedia applications: from ``black and white” videos in old archives and videos with faded colours, to computer vision applications that  discard the chroma component in order to simplify processing. However, while the luminance information provides valuable content-related information regarding shapes and structures, the perception of colour is important for modern video viewing. It is also essential for understanding the visual world, allowing the distinction between objects and physical variations, such as shadow gradations, light source reflections or reflectance variations on video frames. For this reason, adding chromatic information to images and improving the quality of colour has become a research area of significant interest for a wide variety of domains that traditionally have resorted to using luminance data alone. This includes medical imaging \cite{lagodzinski1995colorization}, surveillance systems \cite{haq2010automated} or restoration of degraded historical images \cite{narasimhan2003contrast}. 

Recently, the emergence of deep learning has enabled the development of new colourisation algorithms which better generalise the natural data distribution of colours. Convolutional Neural Networks (CNNs) outperform many state-of-the-art methods based on hand-crafted features in tasks such as image enhancement, image classification or object detection \cite{krizhevsky2012imagenet, lawrence1997face}. State-of-the-art colourisation methods based on Generative Adversarial Neural Networks (GANs) \cite{isola2017image} aim to mimic the natural colour distribution of the training data by forcing the generated samples to be indistinguishable from natural images. Moreover, using adversarial loss, the discriminator can learn a trainable loss function that guarantees a correct adaptation of the differences between  generated and real images in the target domain. However, existing methods still suffer  ambiguity when trying to predict realistic colours often causing desaturated results. Nevertheless,  GANs are a suitable basis for further tackling the desaturation problem and gaining colourfulness.

Motivated by the recent success of Conditional Adversarial Networks in image-to-image translation tasks, including colourisation \cite{isola2017image, wang2018high}, this paper proposes an automatic colourisation paradigm using end-to-end Convolutional Neural Network architectures. Improved colourisation is achieved by introducing techniques that improve the stability of the adversarial loss during training, leading to better colourisation of a variety of images from large multi-class datasets. Further enhancements are achieved by applying feature normalisation techniques which are widely used in style transfer models. The capabilities of adversarial models in image colourisation are improved by adapting an Instance-Batch Normalisation (IBN) convolutional architecture \cite{pan2018two} to conditional GANs. Some examples of the results achieved by the proposed method are presented in Figure~\ref{fig1}. The main contributions of this work are the following:
\begin{enumerate}
  \item Analysis of drawbacks in stat-of-the art methods for automatic image colourisation.
  \item Identify and integrate appropriate architectural features and training procedures which lead to a boosted GAN performance for image colourisation. The proposed steps of improvement include:
  \begin{enumerate}
  \item A novel generator-discriminator setting which adapts the IBN paradigm to an encoder-decoder architecture, enabling generalisation of the content's style changes while encouraging stabilisation during GAN training.
  \item The use of Spectral Normalisation (SN) \cite{miyato2018spectral} for improving the generalisation of the adversarial colourisation and preventing training instability. 
  \item The use of multi-scale discriminators to achieve an improved colour generation in small areas and local details and a boosted colourfulness.
  \end{enumerate}
\end{enumerate}

The paper has the following structure: Section~\ref{background} reviews  related work in the literature, identifying the main drawbacks and possible improvements, Section~\ref{method} details the proposed methodology, Section~\ref{experiments} provides information about the implementation and data used in the experiments and  a quantitative evaluation of the results while  Section~\ref{conclusions} provides  conclusions and identifies  future work.

\section{Background}
\label{background}

Automatic colourisation was originally introduced in 1970 to describe a novel computer-assisted technology for adding colour to black and white movies and TV programs \cite{markle1988coloring}. Although such semi-automatic method improved the efficiency of traditional hand-crafted techniques, it still required a considerable amount of manual effort and artistic experience to achieve acceptable results. Since then, it has been shown that the task is complex, ill-conditioned and inherently ambiguous due to the large degrees of freedom during the assignment of colour information \cite{zhang2016colorful}. 

In some cases, the semantics of the scene and the variations of the luminance intensity can help to infer priors of the colour distribution of the image. For example, an algorithm can successfully associate rapid changes to vegetation areas, assigning ranges of green to it, or smooth areas to sky, inferring blue tones. Nevertheless, in most cases the ambiguity in the decisions can lead a system to make random choices. For instance, the hypothetical prior of a car being red is the same as it being green or blue, although in reality the decision will converge towards the dominant samples in the training data. This fact motivated the research of conservative solutions, such as scribbled-based colourisation \cite{qu2006manga,zhang2017real} and exemplar-based colourisation \cite{bugeau2013variational}, which involve user interaction through semi-automatic methods. In the first method, the user annotates ground truth colours at certain informative points and the system learns how to propagate them to the entire image. Alternatively, a whole colour reference is carefully selected with similar content and semantics to the target, and the system attempts to transfer the colours from reliable estimated correspondences. However, the quality of the final results depends on the choice of the reference samples and the style transfer methodology used to estimate the correspondences.

Another common issue is the well-known desaturated effect \cite{zhang2016colorful,levin2004colorization}, which is associated with treating automatic colourisation as a standard regression problem. Taking a greyscale input image, a parametric model can learn how to predict  corresponding chrominance channels by minimising the Euclidean distance between the estimations and the ground truth. Nevertheless, basic solutions are commonly based on averaging the colours of the training examples. In this way the basic model produces desaturated results characterised by low absolute values in the colour channels when trained on large databases of natural images. Previously this problem has been addressed through a deep learning approach which introduced a rebalancing process during training with the aim of penalising the predicted colours based on their likelihood in the prior distribution of training data \cite{zhang2016colorful}. Such a method outperforms previous state-of-the-art approaches, including recent successes with GANs, in which more complex architectures need to adopt the methodology to generalise the predicted colours.

As proposed in the \textit{pix2pix} framework \cite{isola2017image}, a more traditional regression loss such as $L_{1}$ or $L_{2}$ distance is beneficial when included in the final objective function. This enables a conditional GAN to increase the error rate of the discriminator while producing realistic results close to the  ground truth. Although such a framework achieves state-of-the-art performance across a range of image-to-image translation tasks, it still requires the aforementioned rebalancing method, targeting colour rarity far from the desaturated mean of natural data distributions. The high instability during training when a GAN deals with complex generator architectures and high-resolution training images, can lead the \textit{pix2pix} framework to mode collapse, converging towards  undesirable local equilibria between the generator and discriminator \cite{salimans2016improved,wang2018high}. This effect reduces the contribution of the adversarial loss in the multi-loss objective, giving the total weight of convergence to the regression loss and hence leading the system again to desaturated results.

\begin{figure*}[htbp]
\vspace{-1.5\baselineskip}
\centerline{\includegraphics[height=0.32\textwidth]{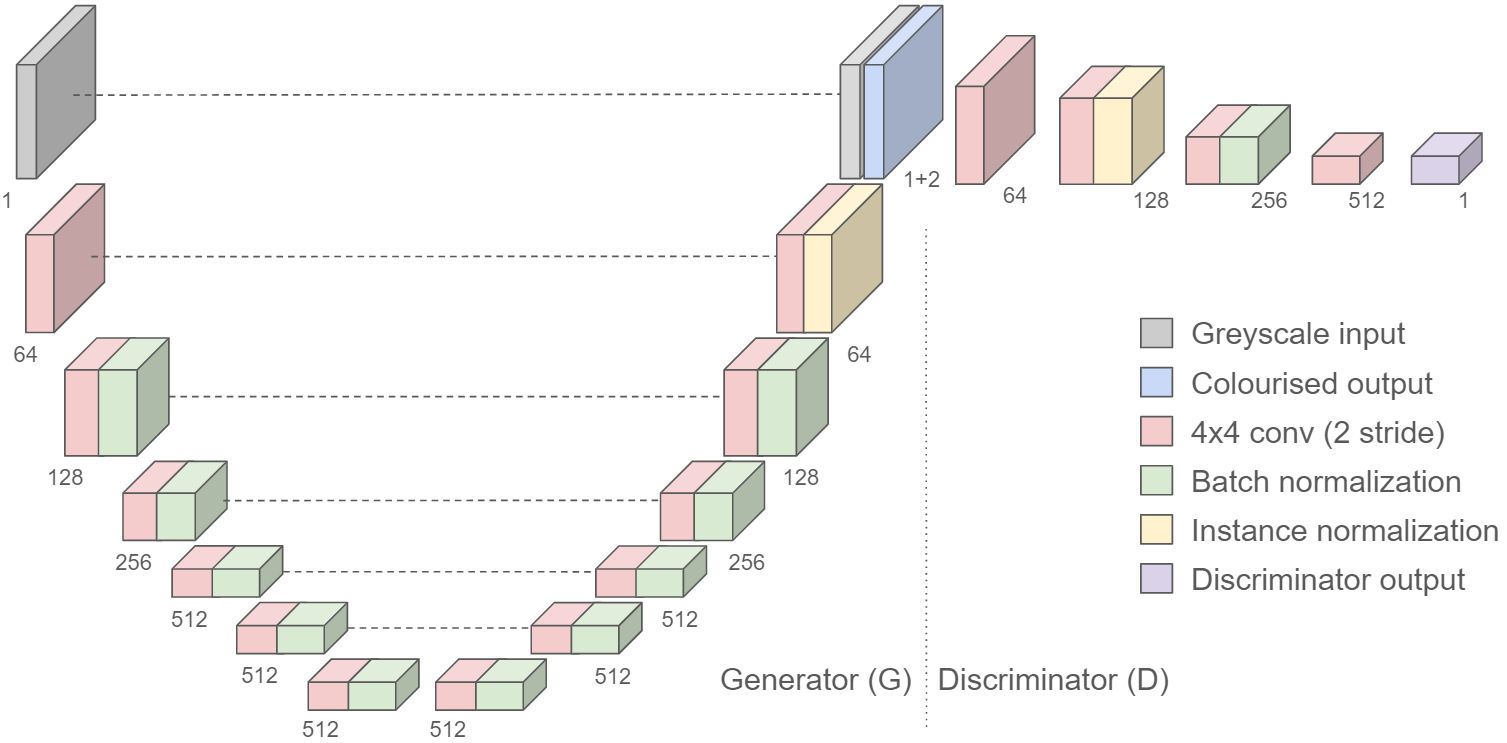}}
\caption{Generator and discriminator architectures with  IBN adaptation to the U-Net and PatchGAN architectures. Note the fake input to the discriminator is composed by the concatenation $(1+2)$ of the original greyscale input and the generated $2$-dimensional colour channels.}
\label{architecture}
\vspace{-0.6\baselineskip}
\end{figure*}

\section{Proposed Method}
\label{method}
Aiming at colourisation of images, the  goal of our  method is to enable automatic CNN-based colourisation of an input greyscale image, denoted $X\in {\rm I\!R}^{H \times W \times 1}$, where $H × W$ is image dimension in pixels, and represented by the lightness channel $L$ in the CIE Lab colour space. To achieve this, it is essential to train an end-to-end CNN architecture capable of learning the direct mapping $\hat{Y} = \mathcal{F}(X)$ to the two associated $ab$ colour channels $Y\in {\rm I\!R}^{H \times W \times 2}$. As commonly used in the literature, CIE Lab colour space is chosen as it is designed to maintain perceptual uniformity and is more perceptually linear than other colour spaces \cite{connolly1997study}. The mapping function $\mathcal{F}(X, \theta)$ can be expressed in a neural network form as:

\begin{equation}
\mathcal{F}(X, \theta)=\mathcal{W}^{L+1}a_{L}(\dots a_{1}(\mathcal{W}^{1}X) \dots ),\label{eq1}
\end{equation}
where $\theta := \{ \mathcal{W}^{1}, \dots, \mathcal{W}^{L+1} \}$ is the set of learning parameters for a $L$-layer CNN, omitting the bias terms for simplicity, and $a_{l}$ the corresponding non-linear activation function, with $l= \{ 1, \dots, L \}$.

\subsection{Conditional Adversarial Networks}\label{cgans}
A mapping convolutional model is trained using a generative adversarial methodology with conditional GANs. This work uses the \textit{pix2pix} framework \cite{isola2017image} as baseline to solve image-to-image translation tasks such as generating realistic street scenes from semantic segmentation maps, aerial photography from cartographic maps or image colourisation from greyscale inputs. As per the traditional GANs setting \cite{goodfellow2014generative}, two CNNs (a generator $G$ and discriminator $D$) are trained simultaneously in a minimax two-player game, with the objective of reaching the \textit{Nash equilibrium} between them. Given an input greyscale image $X$ and a vector of random noise $Z$, the aim of the generator $G$ is to capture the original colour distribution of the training data and to learn a realistic mapping $G(X, Z; \theta_{G})$ to the colourisation result. On the other hand, the discriminator $D$ aims to distinguish real images from colourised ones through the mapping $D(X, Y; \theta_{D})$, estimating the probability that a sample came from the training data rather than from $G$. Therefore, the conditional GAN framework will model the colour distribution of the training data following the minimax training strategy:

\begin{equation}
\min_{G}\max_{D}V(G,D)\label{eq2}
\end{equation}
where the objective function $V(G,D)$ is given by

\begin{equation}
V(G,D) = J^{(G)}(\theta_{G}, \theta_{D}) + J^{(D)}(\theta_{G}, \theta_{D}) \text{, where}\label{eq3}
\end{equation}

\begin{equation}
\begin{split}
J^{(D)}(\theta_{G}, \theta_{D}) = & \EX_{X, Y}[log(D(X, Y))] \ + \\ & \EX_{X, Z}[log(1 - D(G(X, Z)))]  \text{, and}\label{eq4}
\end{split}
\end{equation}

\begin{equation}
\begin{split}
J^{(G)}(\theta_{G}, \theta_{D}) = - & \EX_{X, Z}[log(D(X, G(X, Z))] \  + \\ & \lambda \EX_{X, Y, Z}[\lVert Y - G(X, Y) \lVert_{1}]\label{eq5}
\end{split}
\end{equation}
using $\lambda$ to control the contribution of the regression loss.

As suggested in recent works \cite{goodfellow2016nips,nazeri2018image}, the standard loss function for the generator is redefined in order to guarantee non-saturation by maximising the probability of the discriminator being mistaken and converting the loss to a strictly decreasing function.  Moreover, note the aforementioned $L1$ distance introduced in the final generator objective to encourage a colourisation close to the ground truth outputs. Regarding the GAN architectures, the \textit{pix2pix} framework uses a U-Net \cite{ronneberger2015u} as generator and a Markovian PatchGAN \cite{isola2017image} as discriminator, yielding output probability maps based on the discrimination of $N \times N$ patches in the input domain. They exploit the intrinsic fully convolutional architecture of the discriminator to control the input patch size via its respective receptive field.

\subsection{Mini-batch Normalisation}\label{normalization}
The application of mini-batch normalisation techniques such as Batch Normalisation (BN) \cite{ioffe2015batch}, have become a common practice in deep learning to accelerate the training of deep neural networks. In the case of GANs, the DCGAN architecture it was proven that applying batch normalisation in both generator and discriminator architectures can be very beneficial to stabilise the GAN learning and to prevent a mode collapse due to poor initialisation \cite{radford2015unsupervised}. Internally, batch normalisation preserves content-related information by reducing the covariance shift within a mini-batch during training. It uses the internal mean and variance of the batch to normalise each feature channel.

On the other hand, Instance Normalisation (IN) \cite{ulyanov2016instance} was proven to be beneficial in style transfer speeding-up  fast stylisation. Image colourisation, as other style transfer techniques, aims to capture style information by learning features that are invariant to appearance changes, with the aim to generalise the colourisation process within a mini-batch of variable content. Therefore, unlike batch normalisation, IN uses the statistics of an individual sample instead of the whole mini-batch to normalise features.

Inspired by IBN-Net \cite{pan2018two}, in the presented approach BN and IN are combined in the same convolutional architecture with the aim to exploit the instance-based normalisation capabilities in style transfer while encouraging stabilisation during training, to both improve the learning and generalisation capacities of the GAN. This work adapts the residual IBN-Net architecture to a U-Net generator and a patch-based discriminator. The IBN-Net work discussed that appearance variance in a deep convolutional model mainly lies in shallow layers, while the feature discrimination for content is higher in deep layers. Therefore, IBN-Net avoids IN in deep layers to preserve content discrimination in deep features, while it keeps batch normalisation in the whole architecture to preserve content-related features at different levels. Figure \ref{architecture} shows final proposed architectures for generator and discriminator. Note that normalisation is not applied to the input layers to avoid sample oscillation and model instability.

\subsection{Weight regularisation}\label{regularization}
One common strategy to improve the generalisation of the network and to prevent instability during training is the use weight regularisation. This technique penalises proportionally the weights of the network based on their size, aiming to keep small values during training and hence preventing small changes in the input leading to large changes in the output. In the context of GANs, the use of sigmoid activations in the discriminator can lead the optimisation process towards unbounded gradients. To prevent such a situation, Spectral Normalisation (SN) \cite{miyato2018spectral} was introduced as a regularisation step to control the Lipschitz constant of the discriminator.

In the context of convolutional neural networks, they proved that the Lipschitz constant of a  linear mapping $\mathcal{W}$, \textit{e.g.} between the pre-activations of two layers, is its largest singular value or spectral norm. Then, they performed spectral normalisation by replacing the layer weights $\mathcal{W}$ with $\mathcal{W}/\sigma(\mathcal{W})$,  $\sigma(\mathcal{W})$ being the largest eigenvalue of $\mathcal{W}^{T}\mathcal{W}$. 

\subsection{Multi-scale discriminators}\label{multid}
A challenge in colourisation is to achieve precision in small areas and local details. Using the Markovian PatchGAN discrimination in the \textit{pix2pix} framework, colourfulness can be boosted by increasing the receptive field of the discriminator, albeit at the price of increasing the complexity with deeper architectures and loosing spatial information, commonly leading to blurry effects and tilling artefacts. A better solution is to use the multi-scale discrimination setting to tackle high-resolution image processing without varying the discriminator architecture \cite{wang2018high}. This is achieved using $N$ discriminators at different scales by downsampling the actual inputs. Therefore, keeping fixed the original discriminator architecture, variable receptive fields are obtained. These fields are larger at the coarsest levels, and the modified objective function  $V'(G, D)$ in the GAN context is given as:
\begin{equation}
\begin{split}
V'(G,D) = \sum_{n=0}^{N-1} V(G, D_{n}) \label{eq7a}
\end{split}
\end{equation}
where $\{X_{n}$, $(G(X))_{n} \} \in {\rm I\!R}^{M \times H_{n} \times W_{n} \times C}$ are the $D_{n}$ inputs, with $H_{n} = H/2^{n}$, $W_{n} = W/2^{n}$ and $N$ the number of discriminator scales.

\section{Experiments}
\label{experiments}

\subsection{Data}\label{data}
Training examples are generated from the \textit{ImageNet} dataset \cite{deng2009imagenet}, particularly from the 1,000 synsets selected for the \textit{ImageNet Large Scale Visual Recognition Challenge (ILSVRC) 2012}. Samples are selected from the reduced validation set, containing $50,000$ \textit{RGB} images uniformly distributed as $50$ images per class. The test dataset is created by randomly selecting $10$ images per class from the training set, generating up to $10,000$ examples. All  images are resized to $256 \times 256$ pixels and converted to the CIE Lab colour space.

\subsection{Network architectures}\label{architectures}
The \textit{pix2pix} framework is used as a GAN baseline. The generator consists of a U-Net encoder-decoder architecture, conforming to the following structure:  $e64:e128:e256:e512:e512:e512:e512 - d512:d512:d512:d256:d128:d64$. The encoder's $e$ blocks consist of $4 \times 4$ convolutions with spectral normalisation and a stride of $2$, followed by the normalisation layers as explained in Section~\ref{normalization} and a \textit{Leaky ReLU} activation. The decoder $d$ blocks apply the same block composition but using \textit{ReLU} activations. The last layer is a $4 \times 4$ convolution with a \textit{tanh} activation producing a $2$-dimensional output space. For $L$ layers of encoder-decoder architecture, skip connections between layers $i$ at the encoder and layers $L-i$ at the decoder are applied in order to recover the information lost during the downsampling operations. After the generator, the discriminator is used in a form of $70 \times 70$ PatchGAN with the following fully convolutional architecture: $e64:e128:e256:e512$. The output layer is a $4 \times 4$ convolution producing the output probability maps, and the input layer takes the concatenation between the original greyscale input and the original or generated colour channels. Regarding the multi-scale discrimination, a setup of $3$ different discriminators is used, downsampling the original input volumes by a factor of $2$ and $4$.

\begin{figure}[ht]
\centerline{\includegraphics[width=0.4\textwidth, height=0.5\textwidth]{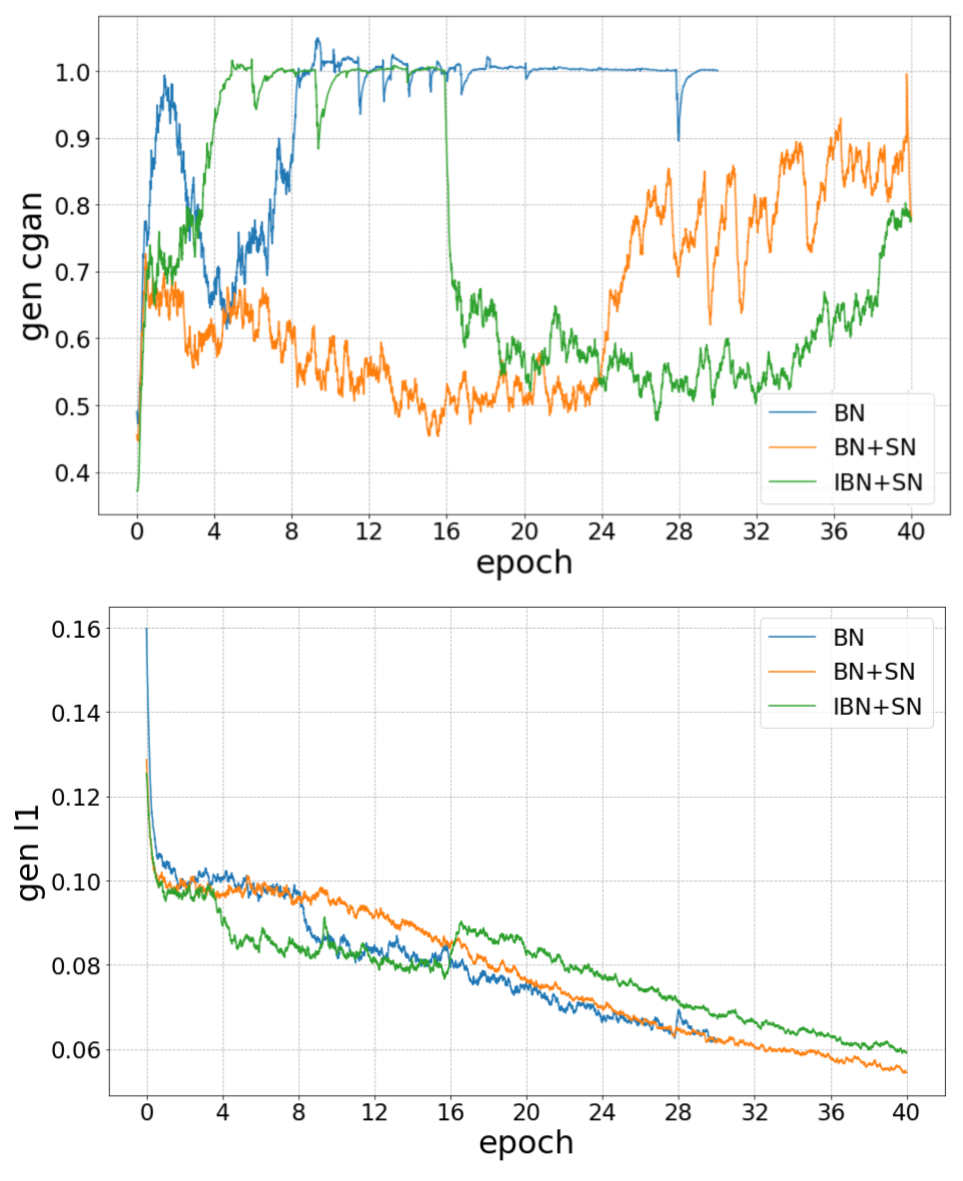}}
\caption{Learning curves of the generator during training. }
\label{curves}
\end{figure}

\begin{figure}[ht]
\vspace{-0.5\baselineskip}
\centerline{\includegraphics[width=0.41\textwidth, height=0.51\textwidth]{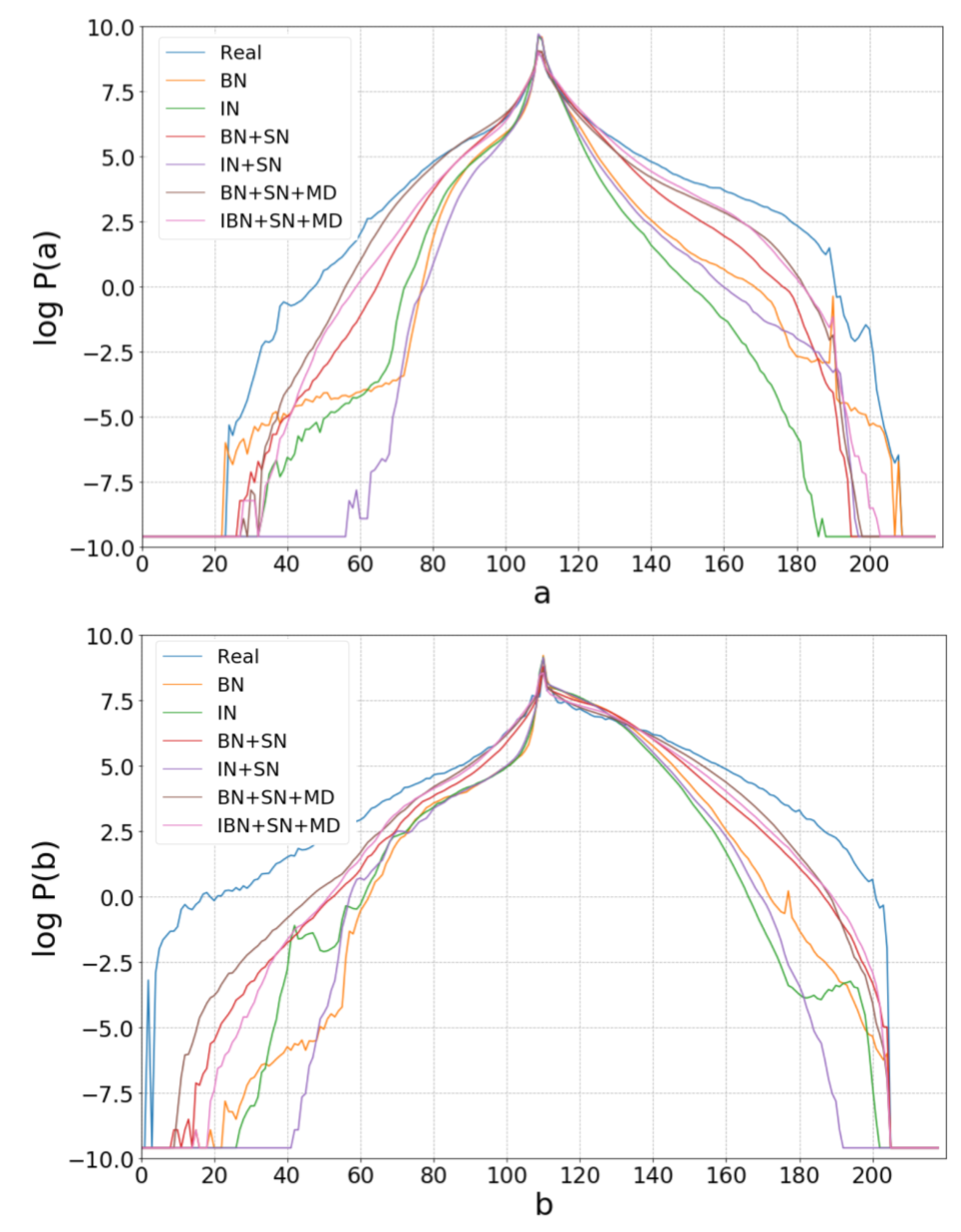}}
\vspace{-0.5\baselineskip}
\caption{Comparison of colour histograms over the test data. }
\label{dist}
\vspace{-1\baselineskip}
\end{figure}

\begin{figure*}[htbp]
\vspace{-2\baselineskip}
\centerline{\includegraphics[height=0.42\textwidth]{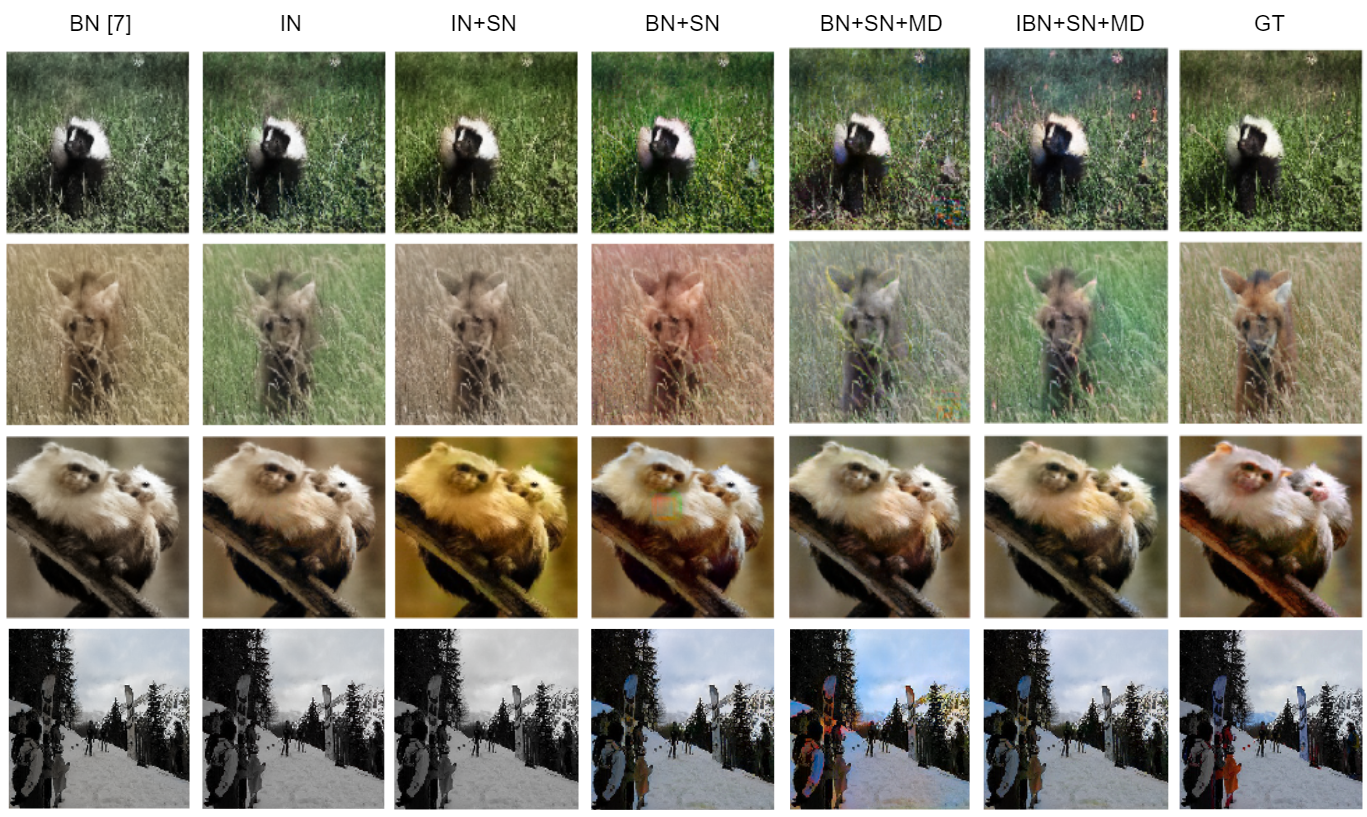}}
\caption{Visual comparison of  colourisation techniques. Note the improvement of the proposed configurations over the BL model: (1) the gain in colourfulness after applying SN, (2) the localisation improvement of MD and (3) the benefits of IBN architecture rather than applying BN and IN separately.}
\label{comparison}
\vspace{-0.32\baselineskip}
\end{figure*}

\subsection{Quantitative evaluation}\label{qresults}
Figure~\ref{curves} illustrates the convergence behaviour of the adversarial and regression losses conforming the generator's objective function. A poor response from  adversarial loss can be observed for the baseline \textit{pix2pix} method, represented by the BN line, which rapidly collapses to a local minimum, giving all the weight of global convergence to the regression loss.  A loss of colourfulness occurs after this point where the regression loss abruptly starts to overfit leading to the generation of desaturated colours. A considerable improvement results after adding spectral normalisation, the BN+SN line, where  weight regularisation helps to stabilise the adversarial loss and  slows down the convergence of the regression, hence preserving colourfulness and preventing overfitting. The aforementioned behaviour can be validated by observing the IBN+SN line. Although instance normalisation leads to instability due to increasing the variance of content-based features during training, a sudden improvement of the adversarial loss can be observed after epoch $16$, where the combination of both normalisation techniques leads to colour generalisation while penalising the regression loss and helping the system to prevent  desaturation. 

The effect of overfitting and lack of colourfulness can be evaluated by comparing deterministic measures, such as the averaged $L_{1}$ or $L_{2}$ distance, with perceptually-based ones designed to better capture the visual plausibility of the results. From the results summarised in Table~\ref{results} it can be observed that, unlike the perceptual evaluation, deterministic measures reflect poor performance for those models generating wider colour distributions, \textit{e.g.} the $L_{2}$ chrominance distance of a red car colourised with a plausible blue will be always higher than being colourised with a desaturated colour. Additionally, the perceptual loss is computed using a \textit{VGG19} model for image classification \cite{simonyan2014very} pretrained on \textit{Imagenet}. As proposed in previous works \cite{johnson2016perceptual,wang2018high}, the $L_{1}$ distance between the convolutional features produced by classifying real and generated samples is averaged as:
\begin{equation}
\begin{split}
\mathcal{L}_{perc} = \dfrac{1}{L} \sum_{i=1}^{L}  \dfrac{1}{N_{i}} \lVert F^{(i)}(x) - F^{(i)}(G(x)) \lVert_{1}^{2} \label{eq7b}
\end{split}
\end{equation}
where $x$ is an input tensor, $F^{(i)}$ are the convolutional features from layers $relu\{i\}\_ 1$, $i \in \{1, \dots, 5 \}$ and $N_{i}$ is the number of features of each volume.

\begin{table}[htbp]
\caption{Quantitative evaluation}
\vspace{-1\baselineskip}
\label{results}
\begin{center}
\begin{tabular}{lccc}
\hline
Method        & \multicolumn{1}{c}{$L_{1}$} & \multicolumn{1}{c}{PSNR {[}dB{]}} & \multicolumn{1}{c}{$\mathcal{L}_{perc}$} \\ \hline
IN            & 9.92              & 26.70                            & 63.85                        \\
BN + SN       & 10.76                      & 25.69                            & 58.58                        \\
IN + SN       & 9.89                       & 26.73                            & 60.36                        \\
BN + SN + MD  & 11.50                      & 25.11                            & 58.52               \\
IBN + SN + MD & 11.20                      & 25.32                          & \textbf{57.77}                        \\
BN (baseline) & \textbf{9.83}                       & \textbf{26.77}                            & 64.05                        \\ \hline
\end{tabular}
\end{center}
\end{table}

Finally, colourfulness is evaluated by estimating the logarithmic colour distribution of the generated $ab$ samples in the test dataset, comparing the proposed configurations with the prior distribution of the real data. As shown in Figure~\ref{dist}, SN provides improved colourfulness for both channels, reducing the area of intersection to the real data distribution with respect to the baseline methodology (BN) with uses Batch Normalisation. Finally, we improved the BN+SN setting by applying  Multi-scale Discrimination (MD), which enables an increase in colourfulness by gaining detail in local and small areas. Examples of colourisation achieved by all analysed methods are presented in Figure ~\ref{comparison}.

\section{Conclusions}
\label{conclusions}
The work presented in this paper improved the state-of-the art for automatic colourisation using conditional adversarial networks. Proposed GAN architecture integrates techniques from the literature to ensure good training stability and to increase the contribution of the adversarial loss during training, which prevents the GAN from collapsing into desaturated colours. It was also shown that batch normalisation and instance normalisation can be integrated together in a fully-convolutional encoder-decoder architecture within a GAN framework without lowering performance, and encouraging the assignation of more plausible colours. Finally, this work shows that by boosting the performance of the adversarial framework, reduction of the desaturation effect can be achieved due to improvement of the discrimination of unreliable colours.


\bibliography{bibliography}
\end{document}